\documentstyle[lcws2002]{article}
\input{psfig}

\def\be{\begin{equation}}
\def\ee{\end{equation}}
\def\bea{\begin{eqnarray}}
\def\eea{\end{eqnarray}}
\def\Re{{\cal R \mskip-4mu \lower.1ex \hbox{\it e}\,}}
\def\Im{{\cal I \mskip-5mu \lower.1ex \hbox{\it m}\,}}

\def\eg{{\it e.g.}}

\def\sub#1{_{\lower.25ex\hbox{$\scriptstyle#1$}}}
\def\tev{\,{\ifmmode\mathrm {TeV}\else TeV\fi}}
\def\gev{\,{\ifmmode\mathrm {GeV}\else GeV\fi}}
\def\mev{\,{\ifmmode\mathrm {MeV}\else MeV\fi}}
\def\mpl{\ifmmode M_{pl}\else $M_{pl}$\fi}
\def\to{\rightarrow}

\def\subw{_{\rm w}}
\def\mh{\ifmmode m\sbl H \else $m\sbl H$\fi}
\def\mch{\ifmmode m_{H^\pm} \else $m_{H^\pm}$\fi}
\def\mt{\ifmmode m_t\else $m_t$\fi}
\def\mc{\ifmmode m_c\else $m_c$\fi}
\def\mz{\ifmmode M_Z\else $M_Z$\fi}
\def\mw{\ifmmode M_W\else $M_W$\fi}
\def\mws{\ifmmode M_W^2 \else $M_W^2$\fi}
\def\mhs{\ifmmode m_H^2 \else $m_H^2$\fi}   
\def\mzs{\ifmmode M_Z^2 \else $M_Z^2$\fi}
\def\mts{\ifmmode m_t^2 \else $m_t^2$\fi}
\def\mcs{\ifmmode m_c^2 \else $m_c^2$\fi}
\def\mchs{\ifmmode m_{H^\pm}^2 \else $m_{H^\pm}^2$\fi}
\def\ztwo{\ifmmode Z_2\else $Z_2$\fi}
\def\zone{\ifmmode Z_1\else $Z_1$\fi}
\def\mtwo{\ifmmode M_2\else $M_2$\fi}
\def\mone{\ifmmode M_1\else $M_1$\fi}
\def\tb{\ifmmode \tan\beta \else $\tan\beta$\fi}
\def\xw{\ifmmode x\subw\else $x\subw$\fi}
\def\ch{\ifmmode H^\pm \else $H^\pm$\fi}
\def\lum{\ifmmode {\cal L}\else ${\cal L}$\fi}
\def\inpb{\,{\ifmmode {\mathrm {pb}}^{-1}\else ${\mathrm {pb}}^{-1}$\fi}}
\def\infb{\,{\ifmmode {\mathrm {fb}}^{-1}\else ${\mathrm {fb}}^{-1}$\fi}}
\def\epem{\ifmmode e^+e^-\else $e^+e^-$\fi}
\def\ppb{\ifmmode \bar pp\else $\bar pp$\fi}
\def\bsg{\ifmmode B\to X_s\gamma\else $B\to X_s\gamma$\fi}
\def\bsll{\ifmmode B\to X_s\ell^+\ell^-\else $B\to X_s\ell^+\ell^-$\fi}
\def\bstt{\ifmmode B\to X_s\tau^+\tau^-\else $B\to X_s\tau^+\tau^-$\fi}
\def\lamt{\ifmmode \tilde\lambda\else $\tilde\lambda$\fi}
\def\shat{\ifmmode \hat s\else $\hat s$\fi}
\def\that{\ifmmode \hat t\else $\hat t$\fi}
\def\uhat{\ifmmode \hat u\else $\hat u$\fi}

\newskip\zatskip \zatskip=0pt plus0pt minus0pt

\begin{document}

\rightline{\vbox{\halign{&#\hfil\cr
SLAC-PUB-10431\cr
May 2004\cr}}}
\vspace{0.8in}

\title{PHENOMENOLOGY OF HIGGSLESS ELECTROWEAK SYMMETRY BREAKING}

\author{ THOMAS G. RIZZO 
\footnote{Work supported by the Department of Energy,  Contract 
DE-AC03-76SF00515}
\footnote{Talk given at LCWS2004, Paris, France, April 2004}
}
\address{Stanford Linear Accelerator Center,
Stanford,\\ CA 94309, USA}

\maketitle\abstracts{
It is possible to construct models based on warped extra dimensions in which
electroweak symmetry breaking takes place without the introduction of any Higgs
fields. This breaking can occur through the judiciuous choice of boundary 
conditions applied to gauge fields living in the bulk. One then finds that 
the fifth components of these bulk fields act as the Goldstone bosons, even 
for the would-be zero modes of the Kaluza-Klein tower. In this talk 
I will discuss the phenomenology of such scenarios, in particular, the 
problems associated with the construction of realistic models due to the 
simultaneous constraints imposed by precision electroweak data, present 
collider search limits and the requirement of perturabtive unitarity in 
$W_L^+W_L^-$ elastic scattering. Future collider signatures for such 
scenarios are also discussed.}

In the SM the conventional Higgs doublet 
plays several roles, in particular, generating the 
fermion as well as the $W/Z$ 
masses with $\rho=1$ and insuring perturbative unitarity (PU) in, \eg, 
$W_L^+W_L^-$ scattering. We can, however, easily imagine electroweak symmetry 
breaking (EWSB) mechanisms wherein things are not quite so simple. 
One of the latest attempts~{\cite {us}} at describing EWSB makes 
use of generalized boundary conditions (BC's) in a 5-$d$ warped, Higgsless 
Left-Right Symmetric model. Such a breaking of gauge symmetries as happens in 
these models cannot occur in the case of the 
usual orbifold BC's due to, \eg,  the periodicity requirement. In addition, 
the usual BC's imposed on the 4-$d$ components of a bulk gauge 
field, $\partial A_\mu |=0$, forces the wavefunction for the lightest mode to 
be flat in the extra dimension and therefore the corresponding 
state to be massless thus leaving all gauge 
symmetries unbroken. Of course the choice of BC's is not arbitrary and must 
be consistent with, \eg, the variation of the action. 

For such a scenario to be successful it must a mechanism to do  
all that the Higgs does without the introduction of additional scalars. 
Fortunately, this scenario indeed gives rise to a pattern of masses 
and couplings for gauge fiels 
which is qualitatively very similar to the usual SM with a doublet 
Higgs.  In this Randall-Sundrum type setup the 
BC's are specifically chosen to break $SU(2)_R\times U(1)_{B-L} \to 
U(1)_Y$ on the Planck brane with the subsequent breaking 
$SU(2)_L\times U(1)_Y\to U(1)_{QED}$
on the TeV brane.  After the Planck scale symmetry breaking occurs, a
global $SU(2)_L\times SU(2)_R$ symmetry remains in the brane picture;
this breaks on the TeV-brane to a diagonal group $SU(2)_D$ 
corresponding to the custodial $SU(2)$ symmetry present in the SM. This 
custodial $SU(2)_D$ helps maintain the tree level $\rho=1$ result.  
We note that in general such a model contains a large number of parameters: 
an overall mass scale, 
the 3 gauge couplings, $g_{L,R,B}$, and also four parameters describing 
the various gauge field kinetic terms, localized on the two 3-branes, which 
we  denote as $\delta_{B,D,L,Y}$. At tree level, two of the gauge couplings 
as well as the mass scale 
are fixed by the values of $G_F$ and $M_{W,Z}$, which we use as input,  
while the remaining ratio, $\kappa=g_R/g_L$, is found to be restricted to 
values not far from unity by various detailed model considerations. In 
pricipal, the brane terms remain unrestricted.

\begin{figure}[htbp]
\centerline{
\psfig{figure=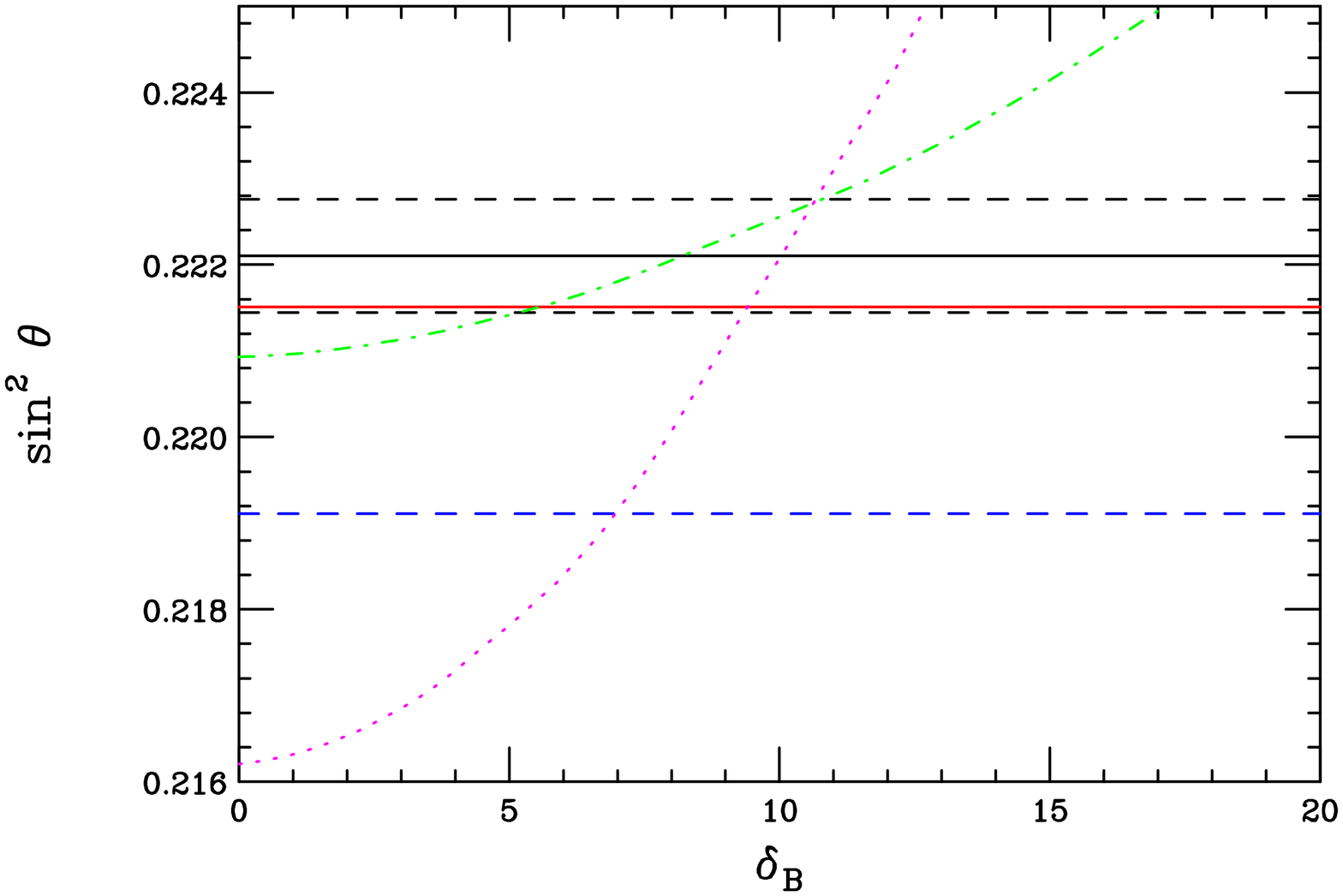,height=5cm,width=5.8cm,angle=90}
\hspace*{1mm}
\psfig{figure=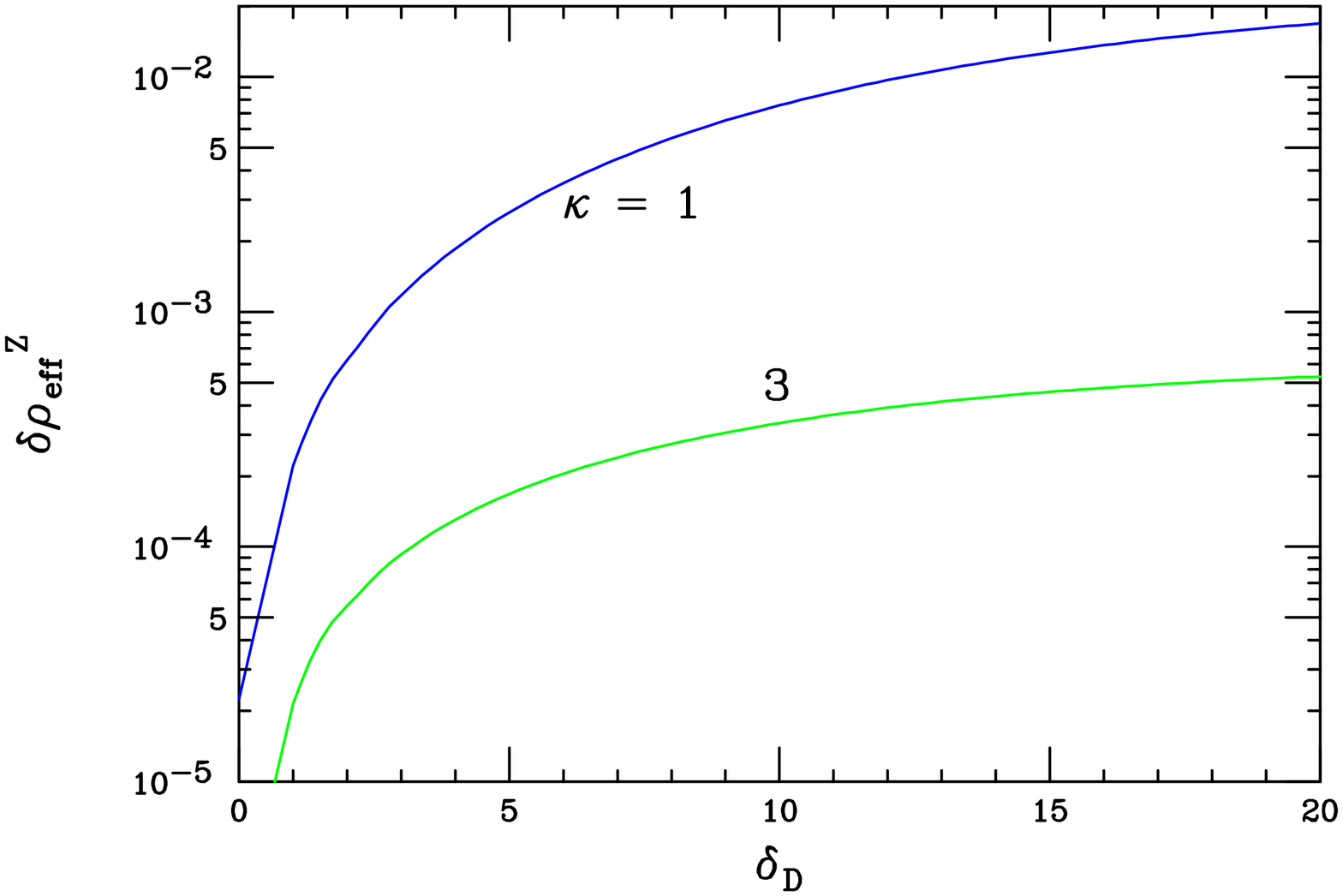,height=5cm,width=5.8cm,angle=90}}
\vspace*{0.1cm}
\caption{(Left) $\sin^2 \theta$ for each of the three definitions as a 
function of $\delta_{B}$.  The black horizontal solid and 
dashed curves correspond to the on-shell value $\pm 1\sigma$,
the solid red (dashed blue) curve represents $\sin^2 \theta_{eff}$ for 
$\kappa=3 (1)$ while the dot dashed green (dotted magenta) curve 
is for $\sin^2 \theta_{eg}$.  We illustrate the effects of including the 
$U(1)_{B-L}$ brane kinetic term. (Right) $\delta \rho_{eff}^Z$ as a function 
of the $SU(2)_D$ brane term $\delta_{D}$ for $\kappa=1$ and 3.}
\label{fig1}
\end{figure}

Unfortunately, as we will discuss below, 
a completely realistic model of this kind has yet to be  
constructed due to the tensions between the various constraints that need to 
be satisfied. 
Not only must the correct pattern of EWSB be obtained but we also demand PU 
while not permitting the gauge boson excitations to be sufficiently light or 
strongly coupled to have shown up at the Tevatron or indirectly 
in contact interaction searches at LEP II. 
Recall that in the SM without a Higgs, PU violation in $W_L^+W_L^-$ 
elastic scattering 
occurs at $\sqrt s \simeq 1.8$ TeV and thus we must expect light 
neutral KK states significantly 
below this mass scale to compensate for the lack of a Higgs. 

An example of one such tension problem in the present scheme  
is the existence of 3 different $\sin^2 \theta$'s  
in this model all of which are identical in the SM at tree level: 
$\sin^2 \theta_{OS}=1-M_W^2/M_Z^2$, which is fixed by the input parameters, 
as well as $\sin^2 \theta_{eg}=e^2/g_W^2$ and 
$\sin^2 \theta_{eff}$ as defined on the 
$Z$ pole. An example of this is shown in Fig.~\ref{fig1}; clearly for a 
successful model we must require that all three of these parameters take on 
very similar values which greatly reduces the size of the allowed parameter 
space. Similarly, we must demand that deviations of the $\rho$ 
parameter from unity as defined, \eg, 
through the $Z$ couplings, must also be small. As we see in another example 
shown in Fig.~\ref{fig1} this too constrains the parameter space as we would 
want $\delta \rho$ to be less than, say, $\sim few ~10^{-3}$. In this 
simple example this would imply that $\delta_D$ not be too large.  It is 
important to observe  
that this set of three quantities; $\delta \rho$ and $\sin^2 \theta_{eg,eff}$ 
can be used to describe all of the deviations from the tree level SM in 
precision measurements. 

\begin{figure}[htbp]
\centerline{
\psfig{figure=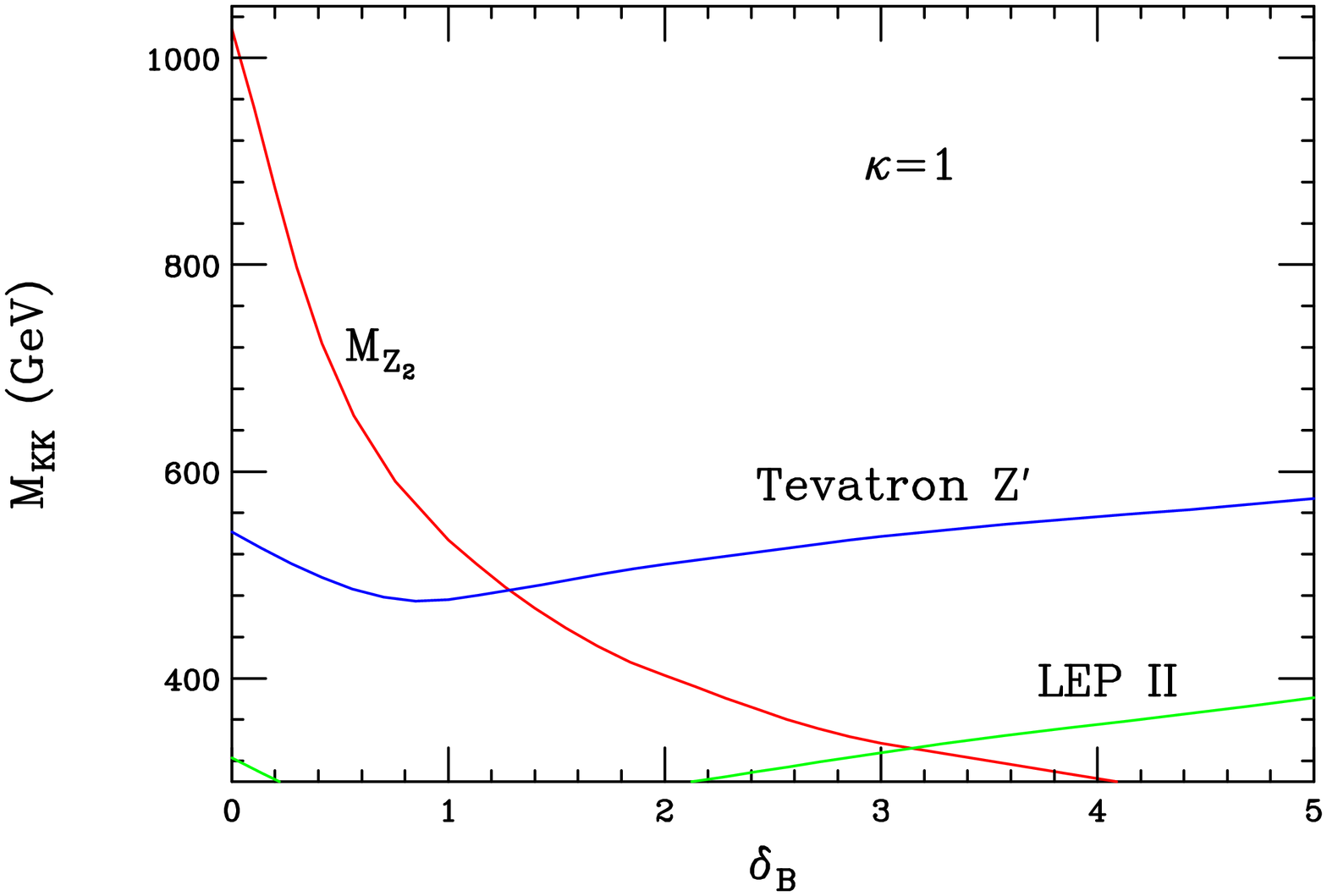,height=5cm,width=5.8cm,angle=90}
\hspace*{1mm}
\psfig{figure=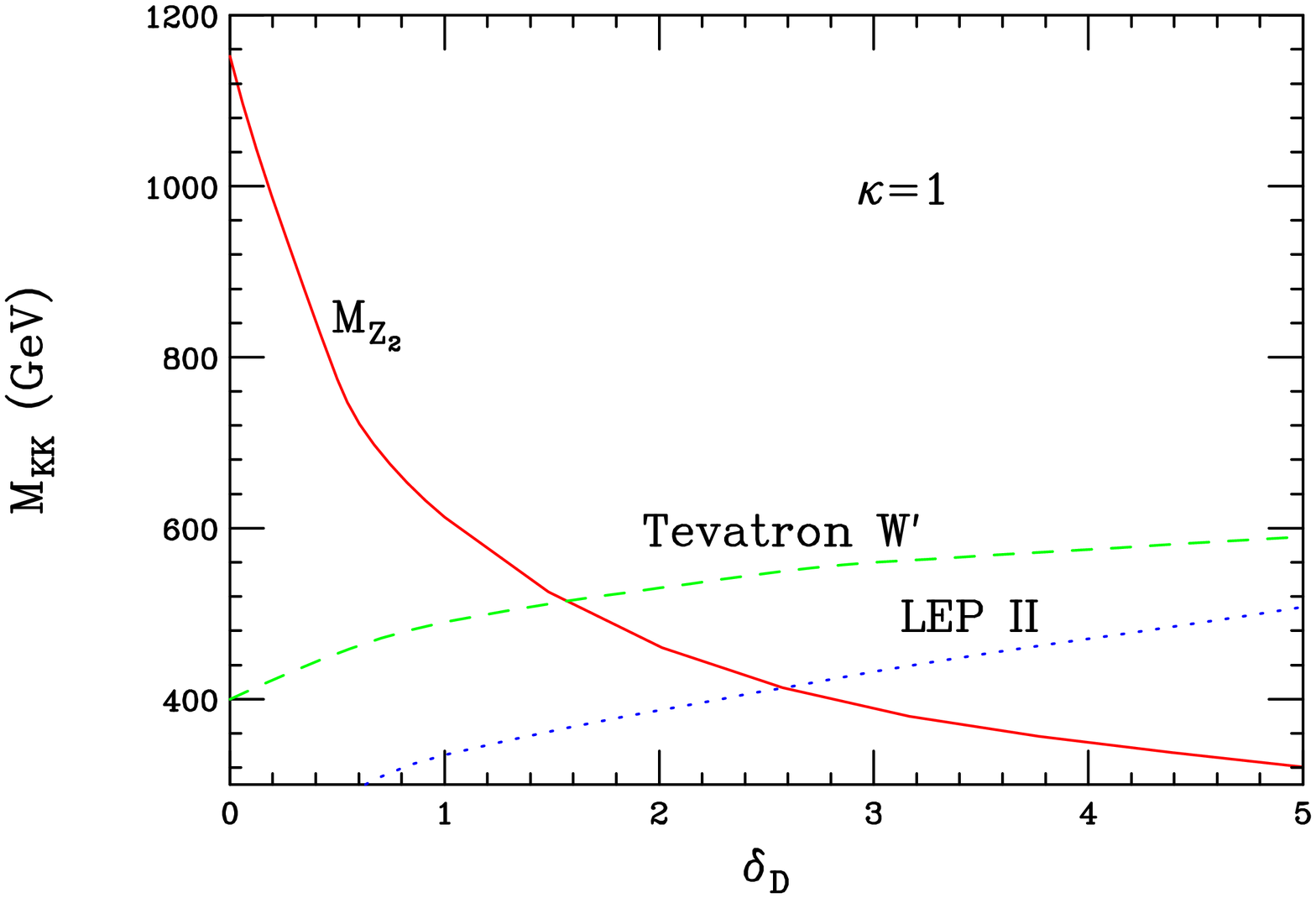,height=5cm,width=5.8cm,angle=90}}
\vspace*{0.1cm}
\caption{(Left) The predicted mass of the lightest KK excitation, the lower 
bound on the mass from the Run II Tevatron $Z'$ searches as well as the lower 
bound from LEPII as a function of $\delta_{B}$; (Right) Same as before but 
now for a non-zero $\delta_D$ and employing the Run I Tevatron bound from $W'$ 
searches.}
\label{fig2}
\end{figure}

The next set of constraints arises from failed searches at the 
Tevatron for the charged and neutral KK excitations, analogous to $W'$ and 
$Z'$ searches, respectively, as well as contact 
interaction bounds from LEP II. Sample constraints on the Higgsless 
model parameter space arising from these considerations are shown in 
Fig.~\ref{fig2}. Here we see that these constraints tend to favor small 
values for the $\delta_i$ parameters corresponding to larger KK masses 
while the `matching' of the three $\sin^2 \theta$'s tend to favor larger 
values for these brane terms.

\begin{figure}[htbp]
\centerline{
\psfig{figure=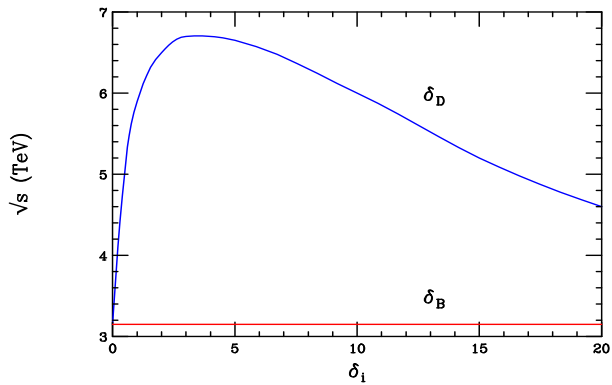,height=5cm,width=8.0cm,angle=0}}
\vspace*{0.1cm}
\caption{The scattering energy at which perturbative unitarity is violated
in $W_L^+W_L^-$ scattering as a function of the kinetic terms.  We
take $\kappa=1$ in this plot.}
\label{fig3}
\end{figure}

A last consideration is PU and its violation in $W_L^+W_L^-$ scattering. In 
the SM the individual diagrams consisting only of gauge bosons each lead to  
amplitudes which grow $\sim s^2$; gauge invariance removes this growth when 
the diagrams are summed yielding a $\sim s$ growth. At this point the 
Higgs contribution enters removing this growth leaving only constant terms 
and results in PU. 
Here, with no Higgs, the $W$ 4-point and $WWZ_n$ couplings must be judiously 
modified to cancel both the $\sim s^2$ and $\sim s$ terms. To explore how well 
this cancellation occurs 
one can ask at what value of $\sqrt s$ PU is   
violated, \eg, 1.8 TeV in the SM with no Higgs, but which 
is essentially infinite in the case 
of the SM with a light Higgs. Clearly, the larger the value of $\sqrt s$ we  
obtain the better we have done at cancelling all of the dangerous terms.  
Fig.~\ref{fig3} shows some sample results for PU violation in the Higgsless 
case. Here we see that 
variations in the brane terms can lead to substantial alterations in the 
scale at which PU violation occurs, for some values of the parameters 
by up to a factor of 4 in 
comparison to the SM with no Higgs. In the most `successful'cases 
the brane term forces the 
lightest neutral KK to couple to isospin thus enhancing its couplings to $WW$. 
Such types of couplings are probably 
necessary in any realistic model in order to 
obtain PU. We have not found, however, any parameter space 
regions where the PU violation scale gets very large, \eg, 100 TeV. 
Some such regions may exist but they have yet to be discovered.

\begin{figure}[htbp]
\centerline{
\psfig{figure=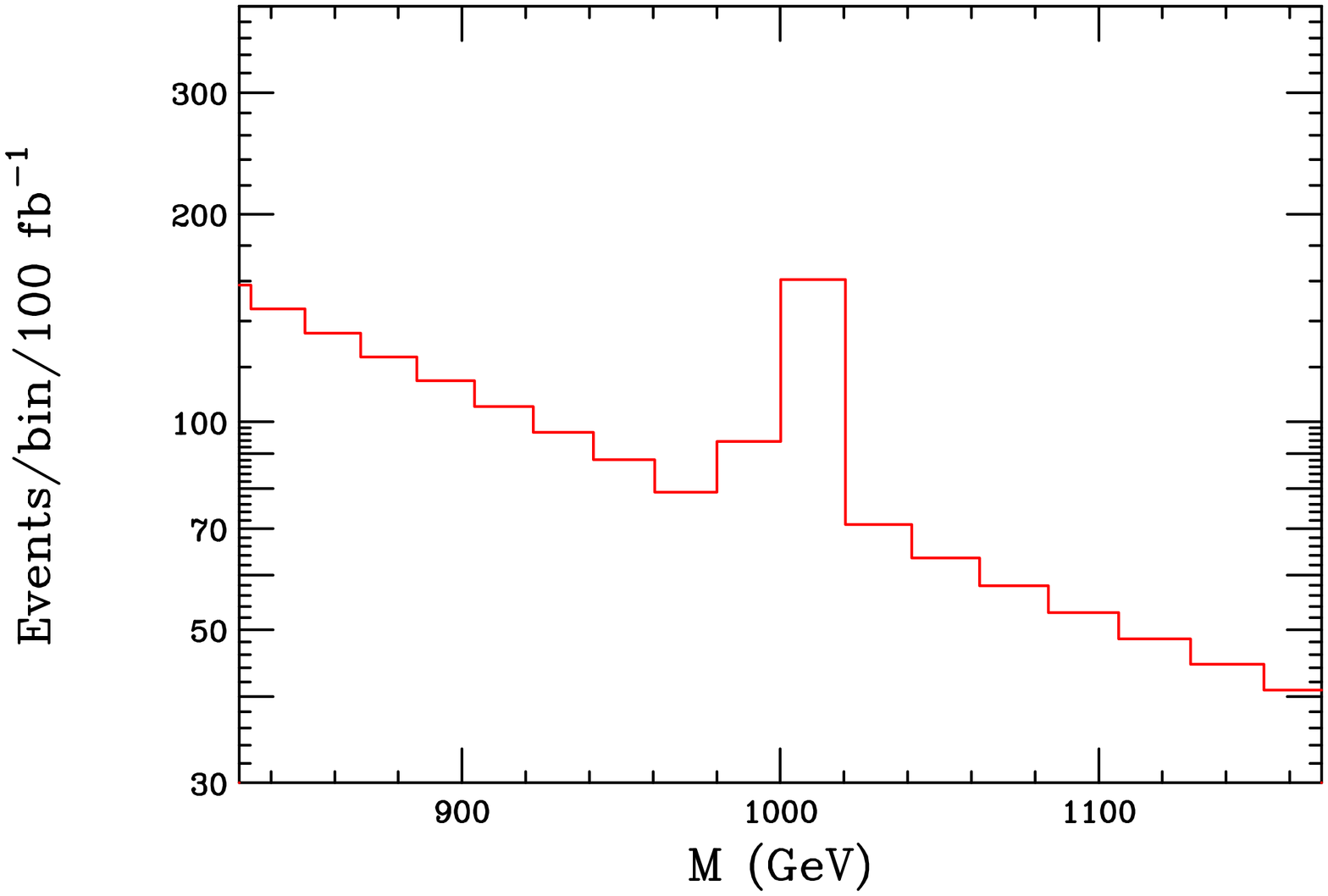,height=5cm,width=5.8cm,angle=90}
\hspace*{1mm}
\psfig{figure=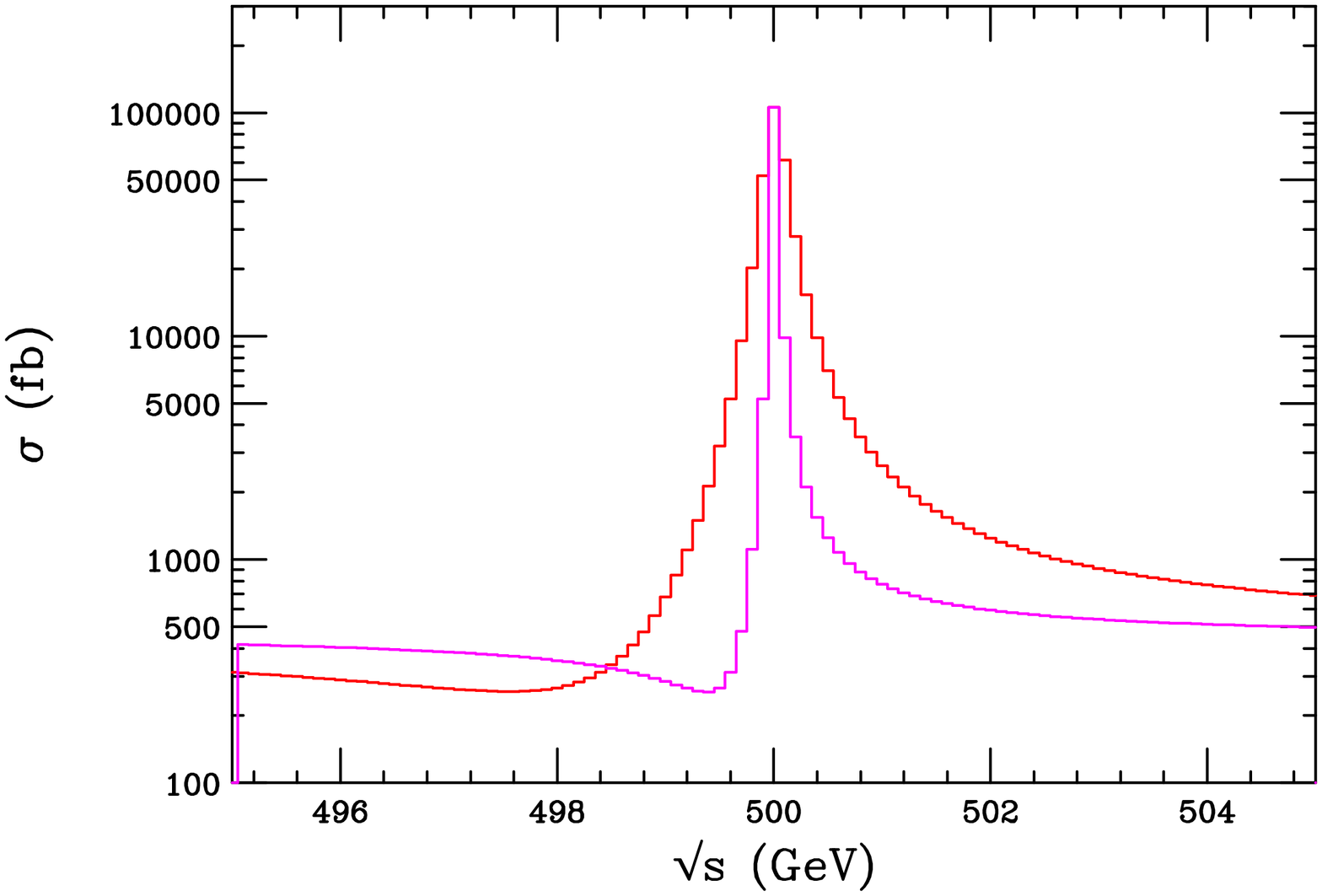,height=5cm,width=5.8cm,angle=90}}
\vspace*{0.1cm}
\caption{(Left) Drell-Yan cross section for a 1 TeV neutral KK coupling 
proportional to isospin with $1/20$ SM strength at the LHC smeared by 
the electron pair ATLAS detector 
resolution. (Right) The corresponding KK unsmeared peaks at the LC for both 
$1/10$ and $1/20$ SM couplings. Smearing is important in both cases due to 
the small width to mass ratio of the KK excitation.}
\label{fig4}
\end{figure}

As can be seen from the discussion and examples 
above it is very difficult for Higgsless 
models to simultaneously satisfy all of the required 
constraints and thus it is not 
trivial to fully imagine what a completely 
realistic model, if it exists, will look 
like. However, it is clear that the existence of light KK excitations 
coupling to isospin will most likely 
be a necessary ingredient if we want to obtain PU. In addition, 
such states must have reduced couplings to the SM fermions on the Planck 
brane in order 
to avoid present search constraints. Thus one should look for light 
KK's at future colliders which are 
somewhat narrow and live in the mass range of 400-1000 GeV. This is an ideal 
match 
for both LHC and LC search capabilities  as can be seen from Fig.~\ref{fig4}.
Since the width to mass ratios of these KK states are expected to be small, 
\eg, $\Gamma/M \sim 10^{-4}-10^{-3}$, detector smearing issues become of 
significance at the LHC as do the corresponding issues of beam energy 
spread at the LC~{\cite {freitas}}.  
It is clear from these figures however that if our qualitative understanding 
of the nature of a `full' theory is correct we can conclude that such KK 
states will be observable at both colliders. This may be necessary as it will  
be the role of the LC to identify the resonance as a KK state arising from 
a Higgsless model once it is discovered at the LHC.

In summary, we have explored the constraints imposed on the construction of a 
successful model of Higgsless EWSB and the possible collider signatures for 
such a scenario. While such a theory does not yet exist, the challenging 
search continues.

\end{document}